\documentclass{article}  
\usepackage{bigsky2007}
\usepackage{graphicx}
\frompage{000} \topage{000}                                              
\def\rightmark{ Color charge dependence of energy loss at RHIC}
\def\leftmark{B. Mohanty}
 
\title{Search for color charge dependence of energy loss at RHIC} 
\authors{{Bedangadas Mohanty$^1$(for the STAR Collaboration) }\\
{\normalsize \hspace*{-8pt}$^1$ Lawrence Berkeley National Laboratory, 
One Cyclotron Road, Berkeley, CA 94720
}
}
 
\abstract{ The non-Abelian feature of quantum chromodynamics (QCD) results in the 
 gluons losing more energy than quarks in the medium formed in high energy heavy-ion collisions. 
Experimental results in p+p collisions when 
compared to NLO pQCD calculations show that at high transverse momentum ($p_{\mathrm T}$)
the produced protons+anti-protons are dominantly 
from gluon jets and charged pions have substantial contribution from quark jets.
If such a scenario is applied to heavy-ion collisions at RHIC, one would 
expect the difference in quark and gluon 
energy loss to have an effect on measured observables, such as high $p_{\mathrm T}$
$\bar{p}(p)/\pi$ ratios and the nuclear modification factor for 
various particles species. We discuss
the experimental results and some possible future measurements.}

 
\begin{document}
 
\maketitle
\setcounter{page}{1}

\section{Introduction}\label{intro}

Results from central heavy-ion collisions at the Relativistic Heavy-Ion Collider (RHIC)
show that at large transverse momentum ($p_{\mathrm T}$ $>$ 6 GeV/$c$) hadron production
is suppressed relative to their production in nucleon-nucleon collisions~\cite{swp}. This is due to 
the medium induced energy loss of the parton propagating through the dense medium 
formed in heavy-ion collisions. Theoretical models incorporating such a mechanism 
find the average energy loss of partons in a static medium
is given as, $<\Delta{E}>$ $\propto$ $\alpha_{s} C_{X} <\hat{q}> L^{2}$~\cite{bdmps}, 
where $\alpha_{s}$ is the strong coupling constant, $C_{X}$ is the color
factor (4/3 for quarks and 3 for gluons), $<\hat{q}>$ is the medium transport co-efficient
which depends on gluon density and parton momenta and $L$ is the path length a parton traverses
in the medium.
In a Bjorken-expanding medium the path length dependence is reduced to 
linear, $ \Delta E  \sim \alpha_{s}^{3} C_X dN^g/dy L / A_\perp$~\cite{glv}, where there are experimental 
handles on the parton rapidity density   $dN^g/dy$ and the transverse area of the collision 
$A_\perp$. A simple estimate of ratio of energy loss of gluons and quarks for similar conditions
in  heavy-ion collisions then is only determined by the color factors of gluon and quark, 
$\Delta{E_{g}}/\Delta{E_{q}}$ $\sim$ 9/4. This indicates
gluons will lose more energy in the medium compared to quarks because of their
stronger coupling. The aim of this paper is to investigate if such a 
difference in energy loss between gluons and quarks can be observed at RHIC.

Before proceeding to present the experimental results we briefly discuss the color factors in QCD.
The color factors are related to the underlying gauge group of QCD and can be estimated
from the dimension of the group. For a SU group of dimension
$N_{c}$, the color factors are given as $C_{A}$ = $N_{c}$, $C_{F}$ = $\frac{N_{c}^{2} - 1}{2N_{c}}$,
and $T_{F}$ = 1/2. Where $A$ and $F$ are the adjoint and fundamental representation of
the group. For QCD, $N_{c}$ = 3, therefore $C_{A}$ = 3 and $C_{F}$ = 4/3. In this theory
the quarks are represented by Dirac fields in the fundamental representation and the
gluons lie in the adjoint representation of SU(3). The color factors are technically the 
eigen values of the Casimir operators of the gauge group and physically they are related to 
the fundamental couplings of the theory. The coupling strength of a gluon to a quark is
proportional to $C_{F}$, $C_{A}$ is related to the strength of gluon
self coupling (a fundamental property of QCD arising due to the non-Abelian nature of
the gauge theory) and the strength of splitting of a gluon into a quark pair is proportional to $T_{F}$ .
The values of the color factors have been experimentally measured at LEP (e.g ALEPH experiment
measures them as $C_{A}$ = 2.93 $\pm$ 0.14 $\pm$ 0.58 and $C_{F}$ = 1.35 $\pm$ 0.07 $\pm$ 0.26 
at $\alpha_{s}$ = 0.119 $\pm$ 0.006 $\pm$ 0.026 )~\cite{lep}. 
The measurements exploited the jet angular distribution 
in events with four jet production in $e^{+}+e^{-}$ collisions. Such events have all three basic 
vertices which are sensitive to the color factors. The measurements experimentally established SU(3) as the
underlying gauge group for QCD. Given their fundamental role in QCD theory and as they appear directly in the 
energy loss calculations which relates to the observed high $p_{T}$ hadron suppression in 
heavy-ion collisions relative to $p$+$p$ collisions, it may be interesting to look for
signatures of effects of color factors on various observables.  
In high-energy collisions, depending on the kinematics, produced hadrons may arise 
from different parton source. In 200 GeV collisions, for example, at mid-rapidity (anti-)protons 
are mainly from gluons while pion production have substantial contribution from quarks.  Therefore these final state
hadrons provide us a powerful tool to test the above mentioned fundamental QCD properties. 
Next we will discuss the experimental results.

\begin{figure}
\vspace{-0.3 cm}
\includegraphics[scale=0.3]{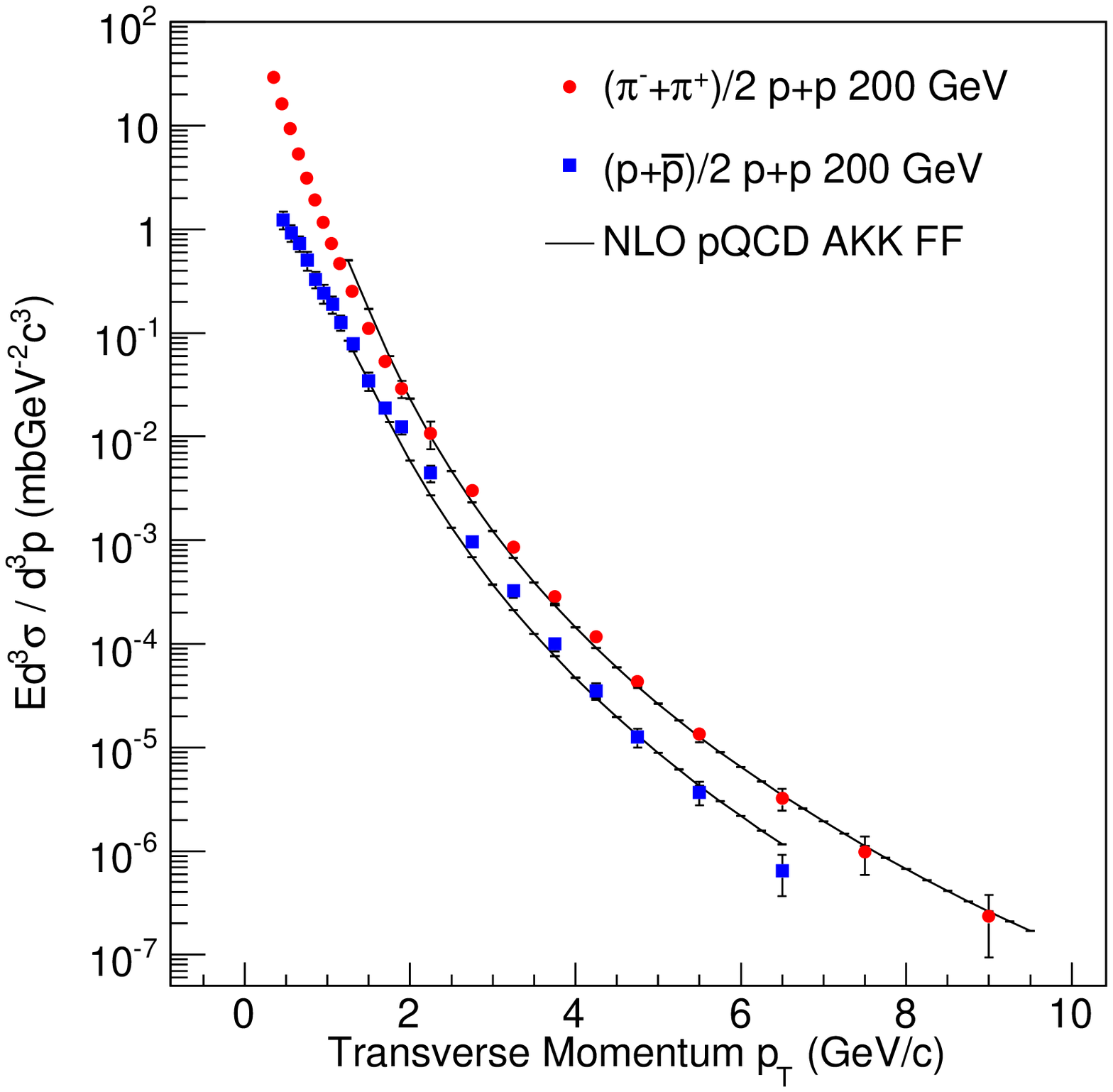}
\includegraphics[scale=0.3]{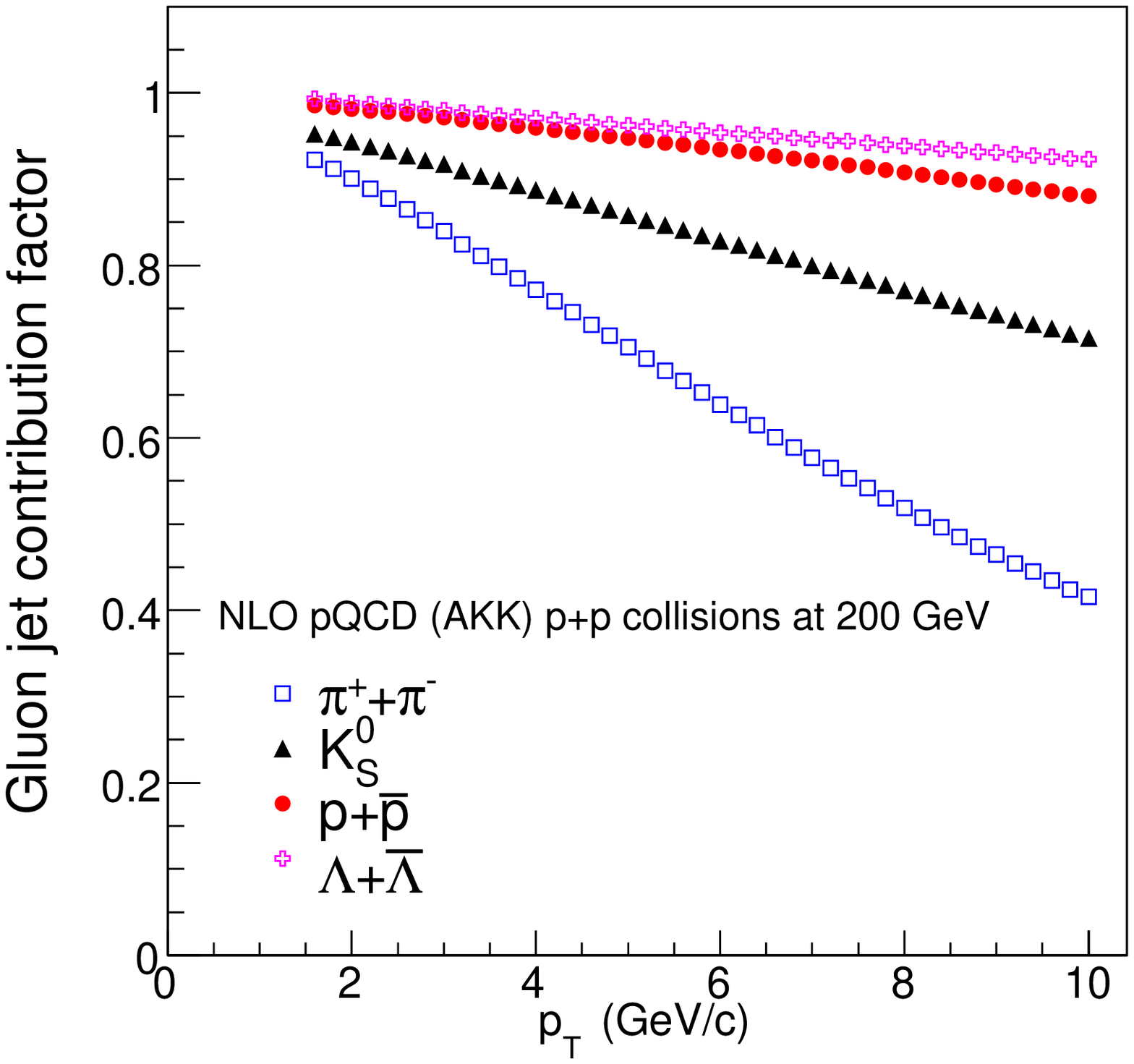}
\vspace{-0.7 cm}
\caption{Left panel : Midrapidity invariant yields for ($\pi^{+}$ + $\pi^{-}$)/2 and ($p$+$\bar{p}$)/2 for
minimum bias $p$+$p$ compared to NLO pQCD calculations using
AKK fragmentation function~\cite{ppdau}. Right panel : Gluon jet contribution factor
to various produced hadrons from NLO pQCD calculations using AKK fragmentation function~\cite{akk}.}
\label{pp_nlo}
\end{figure}

\section{Experimental results}\label{obser}

To study the color charge effect on 
 parton energy loss in heavy-ion collisions we need to focus on the high $p_{T}$ ($>$ 6 GeV/$c$) region 
and identify observables sensitive to quark and gluon jets in heavy ion collisions. 
There are two ways of investigating this effect: (a) at a given beam energy, finding out which of the 
produced hadrons are dominantly coming from quark jets and gluon jets and
(b) at a given $p_{\mathrm T}$, varying the beam energy 
would effectively  mean probing the quark dominated jet production at 
lower beam energy and changing to a gluon dominated jet production at higher energy. 
For example at $p_{T}$ = 10 GeV/$c$, $x_{T}(62.4) \sim 0.32 $ and $x_{T}(200) \sim 0.1 $,
where $x_{T}$ = 2$p_{T}$/$\sqrt{s_{NN}}$, thereby probing regions of different partonic sources. 
NLO pQCD calculations which describe the p+p collisions can be used to get 
an idea about which hadron species are
dominantly produced from quark and gluon jets at RHIC. Figure~\ref{pp_nlo}  shows
that at RHIC the high $p_{T}$ $\pi^{+}+\pi^{-}$ and $p$+$\bar{p}$ 
production is reasonably well described by NLO pQCD calculations
using the AKK fragmentation functions (FF)~\cite{ppdau}. These NLO pQCD calculations
do not provide charge separated results. Figure~\ref{pp_nlo} (right panel) also shows
the gluon jet contribution factor to various produced hadrons as a function
of $p_{T}$ from these NLO pQCD calculations~\cite{akk}. 
The gluon-jet contribution factor is the ratio of contribution to produced
hadron spectra from gluon jets to the total yields from both gluon and quark jets.  
All results are presented for the factorization scale of
$\mu$ = $p_{\mathrm T}$. At high $p_{T}$ the baryons seems
to be dominantly produced from gluon jets ($>$ 90\%), whereas mesons have
significant contribution from quark jets ($\sim$ 20-50\%). This information
drives the choice of various observables discussed below which can be sensitive
to the difference in quark and gluon energy loss.

\subsection{Particle ratios}\label{details} 
\begin{figure}
\vspace{-0.3 cm}
\includegraphics[scale=0.3]{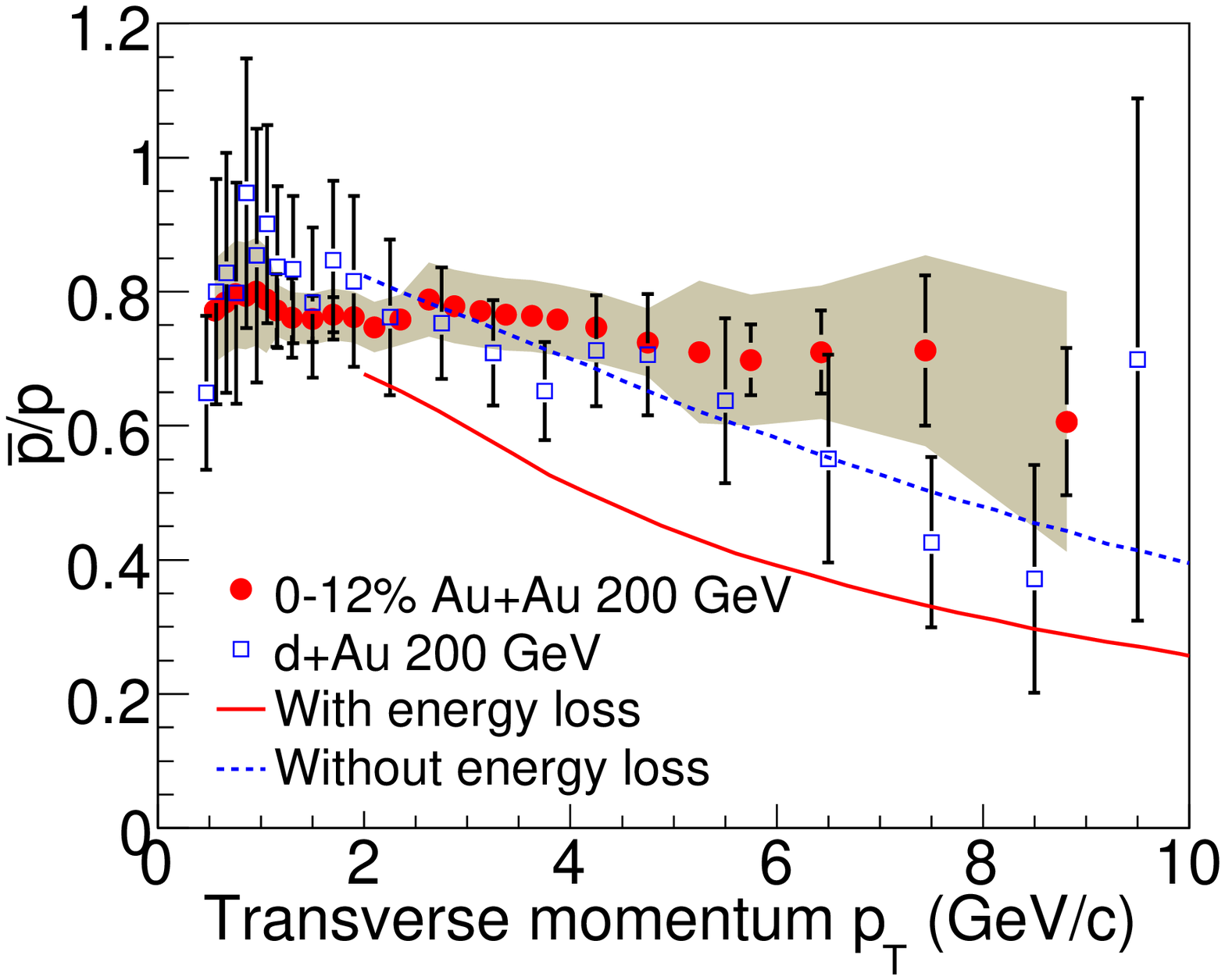}
\includegraphics[scale=0.3]{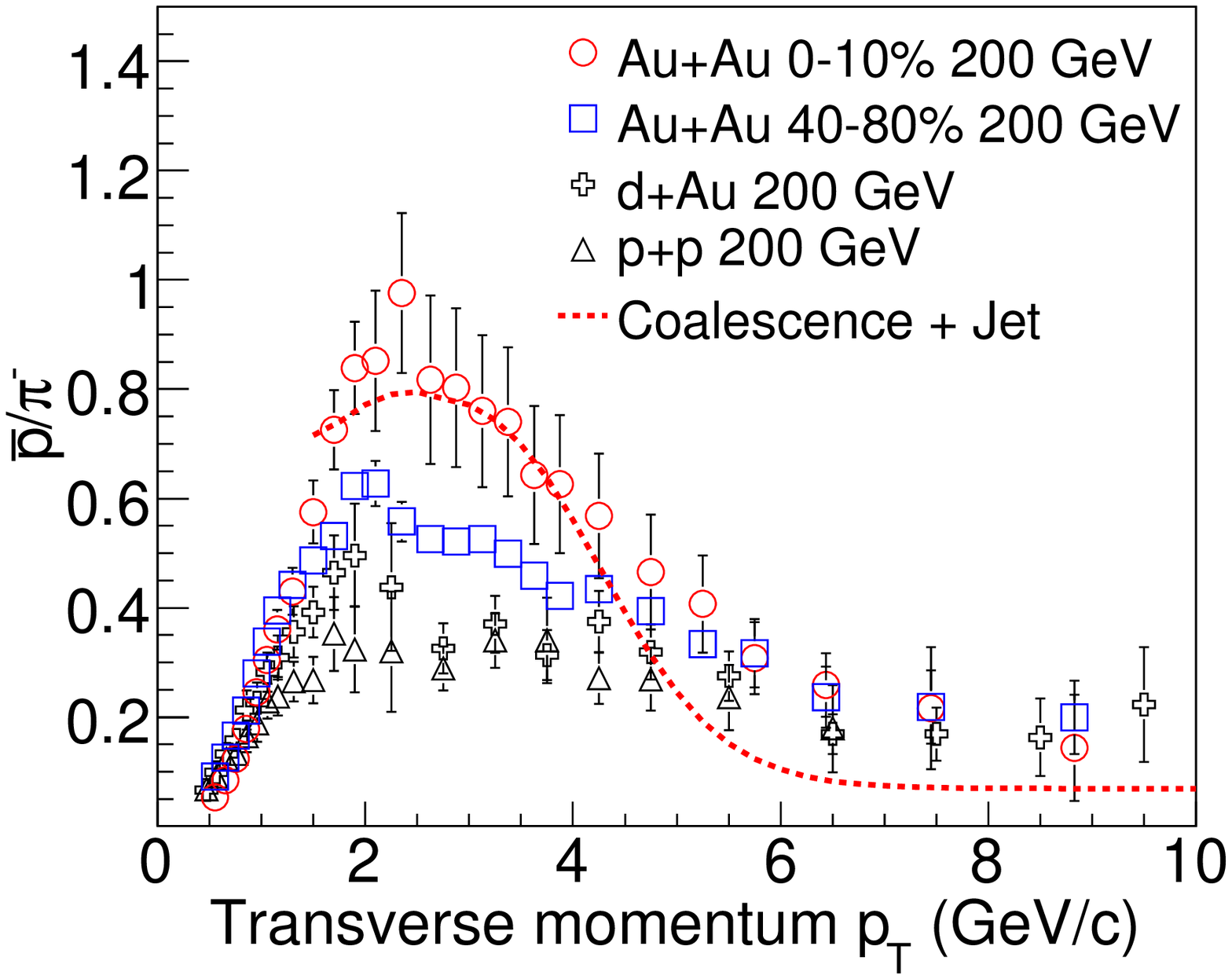}
\vspace{-0.7 cm}
\caption{Left panel : $\bar{p}/p$ ratio vs. $p_{T}$ in central Au+Au and minimum bias $d$+Au collisions
at 200 GeV~\cite{ppdau,auau}. The lines are model calculations with and without energy loss~\cite{wang}. 
Right panel : $\bar{p}/\pi^{-}$ 
ratio vs. $p_{T}$ for central, peripheral Au+Au, minimum bias $d$+Au and $p$+$p$ collisions at 200 GeV~\cite{ppdau,auau}.
Also shown is the calculation from a model based on  
coalescence and jet quenching ~\cite{fries} for central Au+Au collisions.}
\label{aptopratio}
\end{figure}
In quark fragmentation, the leading hadron is more likely 
to be a particle rather than an anti-particle, and there 
is no such preference from a gluon jet~\cite{ppdau}.
If anti-protons are dominantly produced from fragmentation of gluon and protons
have relatively larger contribution from quark jets, then it is expected that for the same 
beam energy, the denser medium formed in central Au+Au collisions will lead to a 
lower $\bar{p}$/$p$ ratio relative to $p$+$p$ or $d$+Au collisions at high $p_{T}$. Similar arguments 
can be made in favour of high $p_{T}$ $\bar{p}$($p$)/$\pi$ ratio. A dense partonic 
medium in central Au+Au collisions where gluons lose more energy than quarks would
results in a lower $\bar{p}$($p$)/$\pi$ ratio at high $p_{\mathrm T}$ 
compared to the corresponding ratios from peripheral Au+Au, $d$+Au or $p+p$ collisions.
Figure~\ref{aptopratio} shows the $\bar{p}/p$ ratio for central Au+Au collisions at 200 GeV 
at high $p_{T}$ ($>$ 6 GeV/$c$) is comparable or slightly higher to $d$+Au results~\cite{ppdau,auau}. 
This is in contrast to the expectations from color charge dependence of energy loss.
Comparison to model calculations~\cite{wang} without energy loss is in reasonable agreement
with the d+Au results, whereas calculations including color charge dependence 
of energy loss give a much lower value of the  $\bar{p}/p$ ratio compared to data 
for most of the measured $p_{T}$ range.
Right panel of figure~\ref{aptopratio} also shows that at high $p_{T}$ ($>$ 6 GeV/$c$)
the $\bar{p}/\pi$ ratios for central, peripheral Au+Au and minimum bias $d$+Au 
and $p+p$ collisions at 200 GeV~\cite{ppdau,auau} are comparable indicating absence of color charge
dependence of parton energy loss. Model calculations  based on  
coalescence and jet quenching ~\cite{fries} (dashed lines) for central Au+Au collisions
predict a much lower value for the ratio at high $p_{T}$.

\subsection{Species dependence of nuclear modification factor}\label{details}
\begin{figure}
\vspace{-0.3 cm}
\includegraphics[scale=0.3]{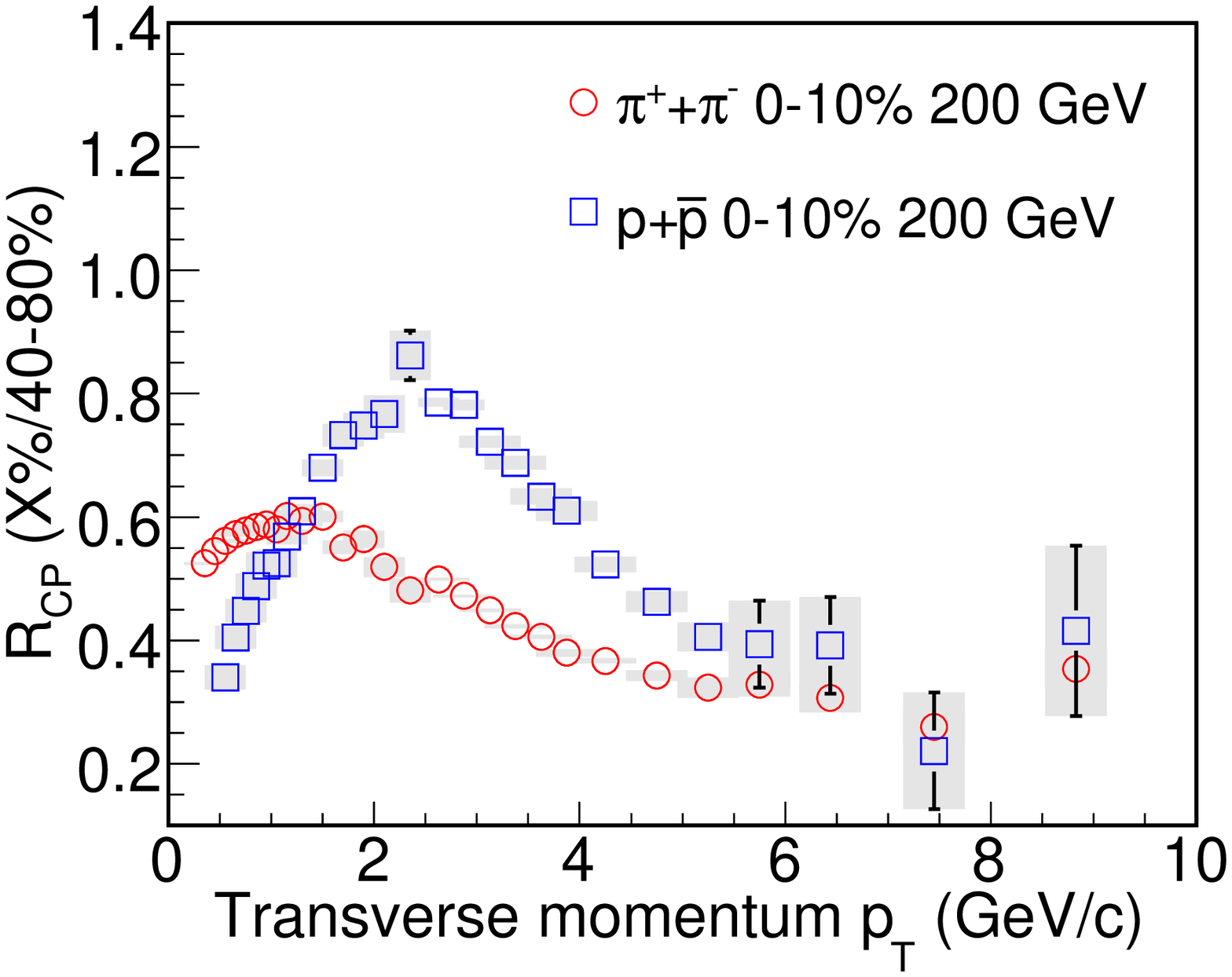}
\includegraphics[scale=0.3]{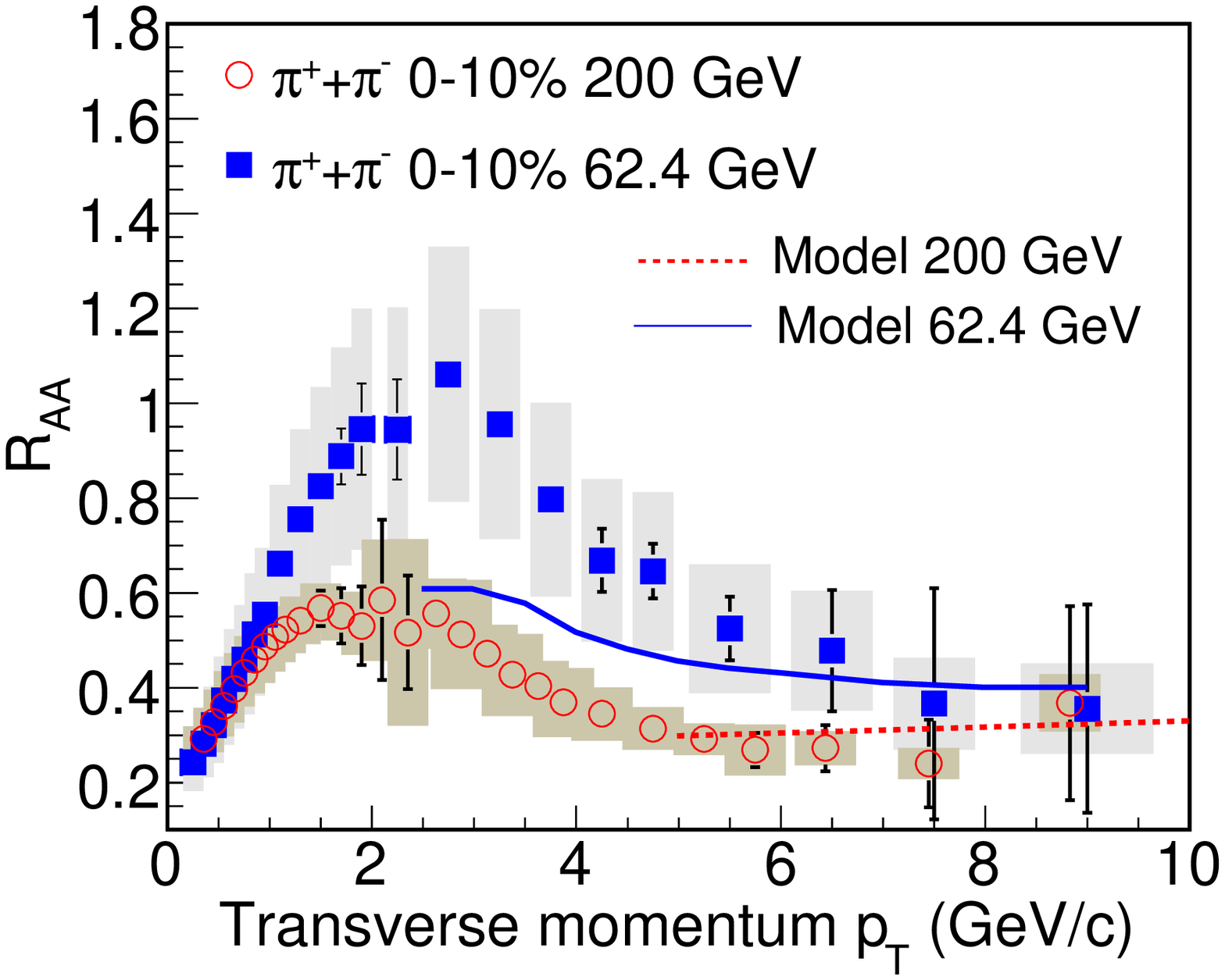}
\vspace{-0.7 cm}
\caption{$R_{CP}$ of $\pi^{+}+\pi^{-}$ and $p+\bar{p}$ and $R_{AA}$ of $\pi^{+}+\pi^{-}$ 
vs. $p_{T}$ in central Au+Au collisions at 200 and 62.4 GeV~\cite{auau}. The lines are model
calculations with energy loss~\cite{vitev}.}
\label{rcp}
\end{figure}

The high $p_{T}$ $p$+$\bar{p}$ production is gluon dominated 
while $\pi^{+}+\pi^{-}$ has significant contribution from quark jets (Fig.~\ref{pp_nlo}). 
The stronger coupling of the gluons with the medium formed in Au+Au collisions will then 
lead to a lower value of the nuclear modification factor (NMF) 
($R_{\rm{CP}}(p_{\rm T})\,=\,\frac{\langle N_{\rm {bin}}^{\rm peri}\rangle d^2N_{\rm{cent}}/dy dp_{\rm T}}{\langle N_{\rm {bin}}^{\rm cent}\rangle \,d^2N_{\rm {peri}}/dy dp_{\rm T}}$)
for $p$+$\bar{p}$ compared  to $\pi^{+}+\pi^{-}$ at high $p_{T}$.
Figure~\ref{rcp} shows the $R_{CP}$ for $p$+$\bar{p}$ is comparable to  $R_{CP}$  of 
$\pi^{+}+\pi^{-}$ at high  $p_{T}$ ($>$ 6 GeV/$c$) for central Au+Au collisions 
at 200 GeV~\cite{auau}. This is in contrast to the naive expectation of
difference in energy loss due to color factors $C_{A}$ and $C_{F}$ being reflected in $R_{CP}$.

\subsection{Energy dependence of nuclear modification factor}\label{maths}

The beam energy dependence of NMF vs. $p_{T}$ provides a chance to probe the
quark dominated jet production and gluon dominated jet production at a given $p_{T}$. Hence in
principle it is sensitive to the color charge effect of parton energy loss. Figure~\ref{rcp}
shows the $R_{AA}$ for charged pions for central Au+Au collisions at 200 and 62.4 GeV~\cite{auau}.
The charged pion $p$+$p$ reference for 200 and 62.4 GeV are from Refs.~\cite{ppdau,david} respectively.
A difference in shape of the high $p_{T}$ dependence of $R_{AA}$ is observed, but
it cannot be attributed to color charge effect without understanding the role
of initial jet spectra and the energy dependence of parton energy loss. Therefore one has to
rely on model comparison. One such model calculation~\cite{vitev}  shown in the figure seems to have a 
reasonable agreement with the measurements at high $p_{T}$. The energy dependence of high $p_{\mathrm T}$
NMF with neutral pions can found in Ref.~\cite{david1}. To get a clear signature of color charge effect 
using this observable, experimental measurements with better precision are need and going to higher beam 
energies will be advantageous.

\section{Discussion on absence of strong color charge effect on energy loss at RHIC}\label{others}

Results from most of experimental observables presented do not indicate any
color charge dependence of parton energy loss in the medium formed in heavy ion collisions
at RHIC. 
We discuss below some of the possible physics reasons for not
observing the color charge effect. 
(a) Can different mechanisms of energy loss (radiative and collisional) smear
    the possibility of observing the difference in energy loss of quarks
    and gluons through hadronic observables ?
(b) Is it because we have gluon dominated matter at RHIC ?
(c) Is there a possibility of quark and gluon jet conversion in the medium
    which leads to an effectively similar quark and gluon energy loss ?
(d) Is the energy of the jet not large enough at RHIC to see the
    difference in fractional energy loss of quarks and gluons ?
(e) Is it possible that there is a two component picture of heavy ion collisions 
    with a core where partons lose all their energy and a corona from where
    the bulk of observed hadrons are emitted ? The high $p_{\mathrm T}$ $R_{AA}$ $<1$ can be 
    then due to absorption of a given fraction of partons in the medium
    (a downward shift in the normalization of the spectra) rather than energy loss of every
    parton (sideward shift in the spectra)~\cite{renk}.
(f) At RHIC $\alpha_{s}$ is fairly large, compared to LEP where the measurements
   of color factors were made. Will a high $\alpha_{s}$ or a stronger coupling
   lead to non-observation of the color effect in energy loss of partons ?
(g) Experimental sensitivity : One set of energy loss calculations with FF which
    describe RHIC data and using a hydrodynamical description of the soft matter
    evolution, shows a difference between pion and $p+\bar{p}$ $R_{AA}$ due to
    color charge effect. However the difference cannot be resolved given the present
    uncertainties in the measurements~\cite{renk}.

{\it Different energy loss mechanism at RHIC :} At RHIC we observe a similar 
suppression pattern at high $p_{T}$ in Au+Au relative to $p$+$p$ collisions 
for particles consisting of light flavour quarks ($\pi$, $\eta$ and $p+\bar{p}$) 
and heavy quarks (non-photonic electrons from semi-leptonic decay of heavy quark mesons)~\cite{nonph}. 
This led to the possibility of significant contribution to energy loss of partons 
through collisional process in addition to radiative energy loss.  
We discussed that the radiative energy loss is directly 
proportional to the color factor. Recent calculations ~\cite{wicks} have found 
similar dependence for the collisional energy loss. The parton nuclear modification
factor with both radiative and collisional energy loss included shows a significantly
larger difference between gluon and quark energy loss. This rules out
the possibility of different energy loss mechanisms being responsible for the 
absence of the color charge effect in the experimental observables presented.
\begin{figure}
\vspace{-0.3 cm}
\includegraphics[scale=0.26]{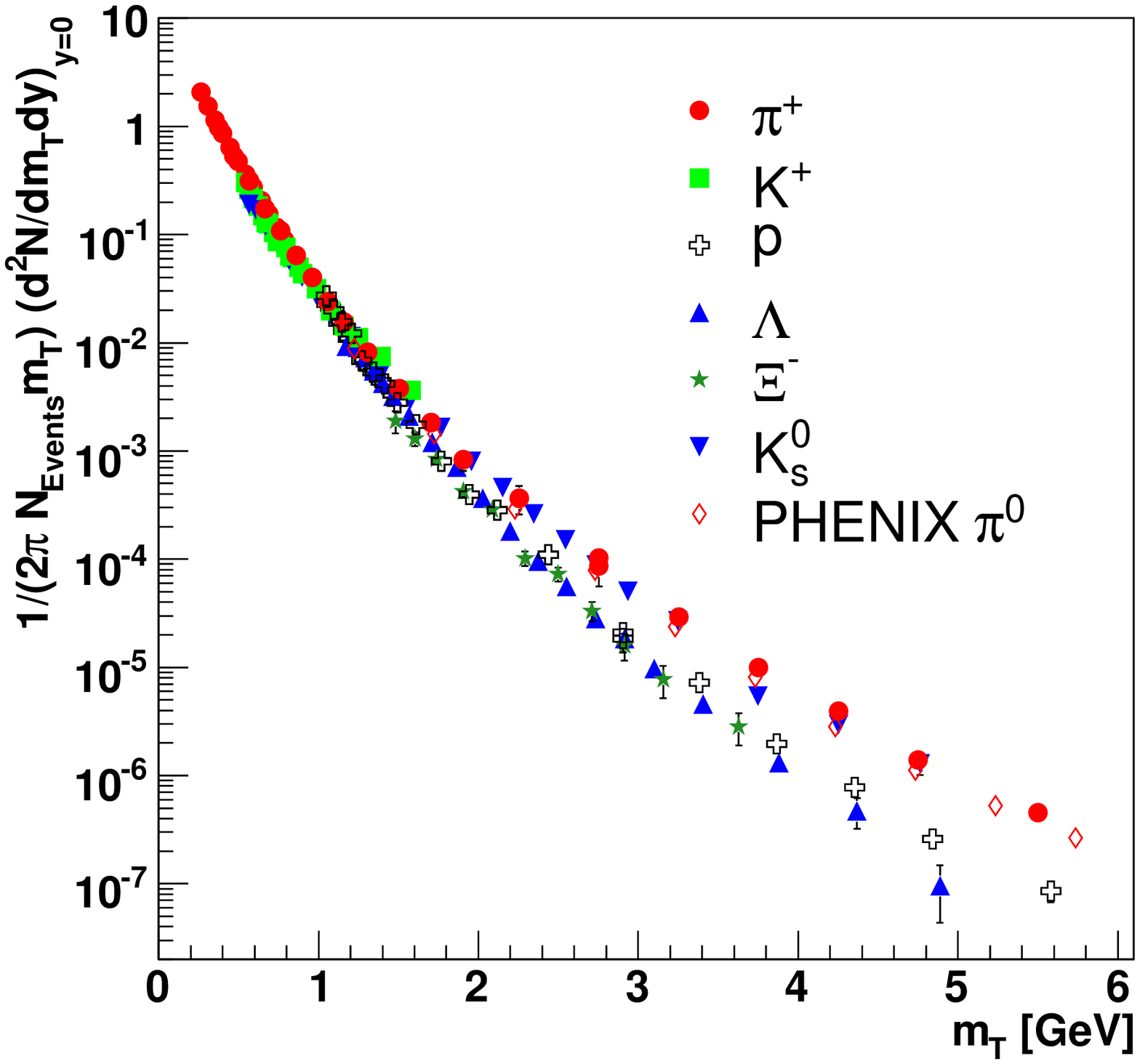}
\includegraphics[scale=0.28]{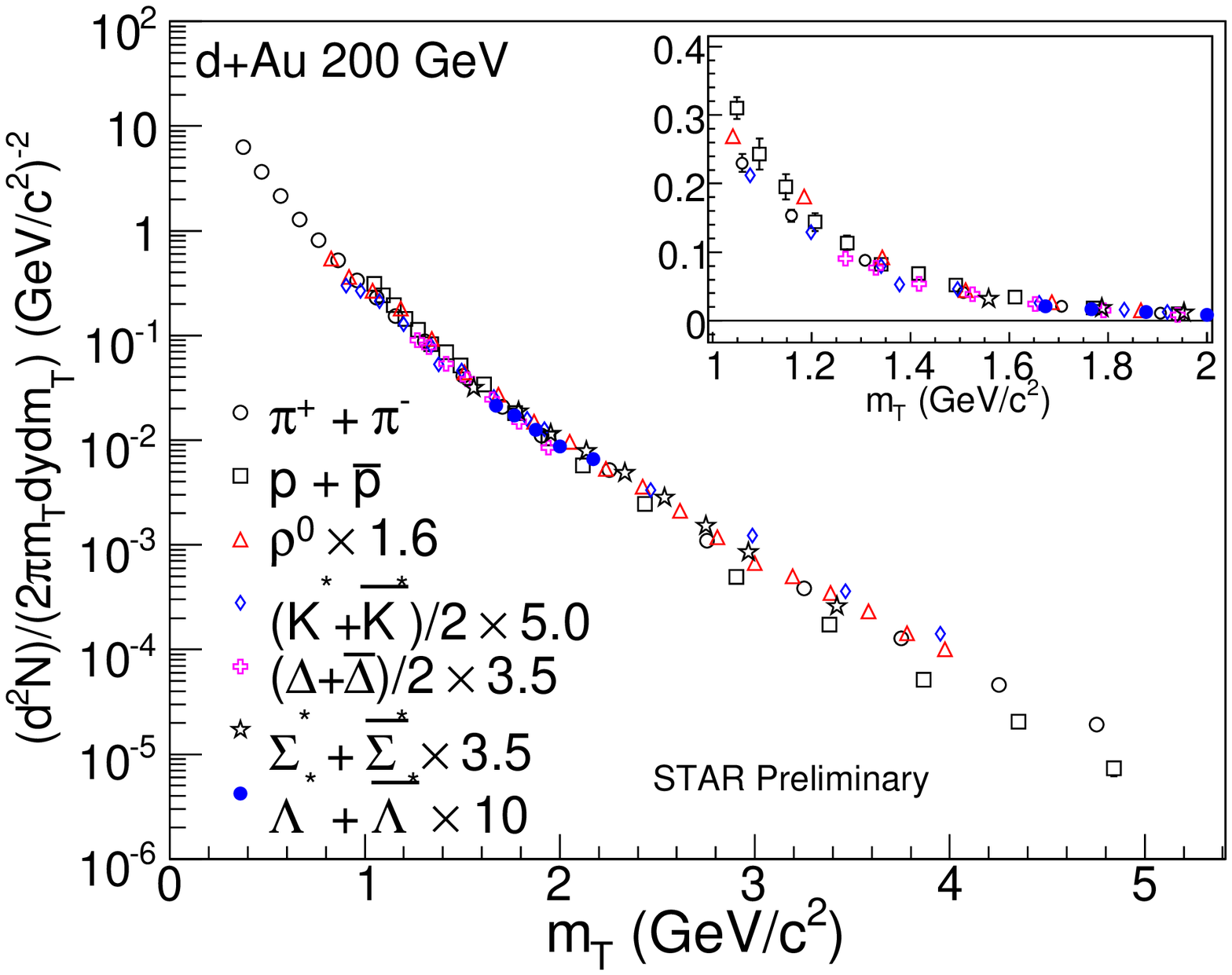}
\vspace{-0.7 cm}
\caption{Scaled transverse mass mid rapidity spectra for various hadrons
from $p$+$p$ (left)~\cite{ppdau} and $d$+Au (right) collisions at 200 GeV.
}
\label{pp_mt}
\end{figure}

{\it Gluon dominated matter at RHIC :} At RHIC we observe a splitting along 
baryon and meson lines in the produced hadron $m_{T}$ spectra at high $m_{T}$ 
(Figure~\ref{pp_mt})~\cite{ppdau}. Pythia simulations show this is a characteristic feature 
of gluon jet events. For quark jet events, simulations show a splitting based 
on mass of the hadron. Preliminary analysis of STAR experiment $d$+Au collision 
data (Figure~\ref{pp_mt}) shows a generalized $m_{T}$ scaling for various particles 
at low $p_{\mathrm T}$. 
Such a scaling is predicted from a color glass condensate scenario in the initial state~\cite{cgc}. 
Furthermore, the $(p+\bar{p})/(\pi^{+}+\pi^{-})$ ratio from quark jets in $e^{+}+e^{-}$
collisions are lower compared to those from Au+Au collisions~\cite{auau}. 
All these observations 
taken together indicate that for the $p_{\mathrm T}$ range studied, there may be 
significantly large contributions from gluon jets to all particle production at RHIC.
So there is a need to carry out  measurements at still higher $p_{T}$ to see a clear 
effect of color charge on various observables.

{\it Quark and Gluon jet conversions :} Recently a theoretical~\cite{ko} attempt has been made 
to understand why the observed $p(\bar{p})/\pi$ at high $p_{T}$ in Au+Au and $d$+Au 
collisions do not reflect the color charge effect of the energy loss. One way by which 
the effect of color charge gets reduced is by allowing for conversions 
between quark and gluon jets through both inelastic ($q\bar{q}$ $\leftrightarrow$ $gg$) and elastic 
($gq(\bar{q})$ $\rightarrow$ $q(\bar{q})g$) scatterings with thermal partons in a
quark gluon plasma (QGP). The conversion rate depends on the collisional width and
is found to be larger for quark jets than gluon jets for a
chemically equilibrated QGP. This increases the final abundance of gluon 
jets and hence compensating for their larger energy loss in QGP. In such 
a scenario it is observed that if the net quark to gluon jet conversion rate 
in heavy-ion collisions is much larger (collisional width enhanced by $\sim$ 6 times) 
than that given by the lowest order in QCD, the results are in reasonable agreement with experimental 
high $p_{T}$ $p/\pi$ ratio~\cite{ko}.

{\it Jet energy and energy loss fluctuations :}
Recent theoretical calculations~\cite{vitev} based on the Guylassy-Levai-Vitev (GLV) approach 
to the medium-induced non-Abelian energy loss with realistic probability
distribution of energy loss and multi gluon fluctuations indicate that 
only in the limit $E_{jet}$ $\rightarrow$ $\infty$ and $\Delta E$/E $\rightarrow$ 0 
does the energy loss for quarks and gluons approach the naive ratio 
$\Delta E_{g}/\Delta E_{q}$ = $C_{A}$/$C_{F}$=9/4. For large fractional energy 
losses this ratio is determined by the $\Delta E$ $<$ E constraint, thereby indicating
we need to move to higher momentum region to see the color charge effect.

{\it $\alpha_{s}$ value at RHIC and accurate determination of color factor :}
Experimental analysis for measuring the color charge factors using O($\alpha^{3}_{s}$)
QCD predictions instead of O($\alpha^{2}_{s}$) results in an introduction of a 
relative factor of about 1 + 2 $\alpha_{s}$ in the ratio of the color factors
$T_{F}$/$C_{F}$~\cite{alphas}.  This together with the  observation of the charged particle
multiplicity ratio of gluon to quark jets (which should be equal to $C_{A}/C_{F}$~\cite{mult}) 
approaching the $C_{A}/C_{F}$ value asymptotically with increasing $Q^{2}$~\cite{tom} 
indicates the possibility
of color charge being increasingly screened at low $Q^{2}$ or larger $\alpha_{s}$ 
values (RHIC : $\alpha_{s}$ values considered is around 0.2 - 0.4). So one possible
question can be, within the typical $Q^{2}$ values encountered within the medium
formed in heavy ion collisions at RHIC, is it possible to have detectable
color charge differences ? In other words is it possible to resolve the different
color charge carriers at relatively small $Q^{2}$ or large $\alpha_{s}$ values ? These possibilities 
indicate having measurements at higher momentum than currently available may be neccessary to
see the color charge effect. 

\section{Possible future measurements}\label{others}
In view of no clear evidence of color charge effect on energy loss observed at RHIC,
it may be worthwhile to discuss of more promising future measurements in addition to
extending the current  measurements to higher $p_{\mathrm T}$ and higher beam energy.
The ratio of NMF of high $p_{T}$ heavy-flavored mesons to 
light-flavored mesons ($R_{D/h}$) in heavy ion collisions can be sensitive to color charge
dependences of medium-induced parton energy loss~\cite{predic}. This ratio is affected by 
(a) mass dependence of parton energy loss, (b) difference
in partonic $p_{T}$ spectrum for light and heavy quarks, (c) difference in light
and heavy quark fragmentation function and (c) color charge dependence of 
parton energy loss, where charm mesons and beauty mesons exclusively probe
the quark energy loss in the medium. However for $p_{T}$ $>$ 14 GeV/$c$ the
ratio $R_{D/h}$ being $>$ 1 is solely due to color charge effect on parton energy loss. This can be
a very clean signature for observing the color charge effect. On similar lines,
the ratio of NMF of high $p_{T}$ $\phi$ meson to 
light-flavored mesons in heavy ion collisions can also be sensitive to color charge
effect of parton energy loss, as the $\phi$ meson is dominantly formed by coalescence 
of s-quarks~\cite{phi}. Looking for difference in the species dependence (pions and anti-protons) of 
suppression pattern in away side ($\Delta \phi$ $\sim$ $\pi$) identified particle 
di-hadron correlation can
also be considered as a signature of color charge effect on parton energy loss. 

\section{Summary}\label{concl}
The non-Abelian features of QCD suggest that gluons, which have a stronger coupling than quarks
with the medium formed in heavy-ion collisions, lose more energy. Observation of this effect 
will link the experimental observations in high
energy heavy ion collisions to one of the basic ingredients
of QCD, the gauge group. So far all the measurements
at high $p_{T}$ believed to be sensitive to color charge effect on medium induced 
partonic energy loss like, $\bar{p}$/$p$, $\bar{p}$/$\pi^{-}$,
$R_{CP}$ of $\pi^{+}+\pi^{-}$ and $p+\bar{p}$  do not show
the naively expected results due to difference in quark and gluon energy loss. We have discussed
some of the possibilities that can lead to an absence of this effect.
In addition to extending the measurements to high $p_{T}$ and higher beam energy, 
measurements of the ratio of nuclear modification factor of high $p_{T}$ heavy-flavored mesons to 
light-flavored mesons and $\phi$ meson to light-flavored mesons in heavy ion collisions
or studying the species dependence of suppression pattern in away side ($\Delta \phi$ $\sim$ $\pi$) 
of identified particle 
di-hadron correlation may be considered as an alternative way of investigating the 
color charge effect on medium induced parton energy loss at RHIC.

\vfill\eject
\end{document}